\documentclass[twocolumn]{aastex62}

\usepackage{amsmath,amssymb,amstext}
\usepackage{natbib}

\newcommand{\um}{$\mu$m}

\shorttitle{Lensed Red Quasar}
\shortauthors{Glikman et al.}

\begin{document}

\title{A highly magnified gravitationally lensed red quasar at $z=2.5$ with significant flux anomaly: Uncovering a missing population}


\author[0000-0003-0489-3750]{Eilat Glikman}
\affiliation{Department of Physics, Middlebury College, Middlebury, VT 05753, USA }
\affiliation{Visiting astronomer, Institute for Astronomy, University of Hawaii, Honolulu, HI, USA }

\author[0000-0003-4561-4017]{Cristian E. Rusu}
\affiliation{Subaru Fellow, Subaru Telescope, National Astronomical Observatory of Japan, 650 N Aohoku Pl, Hilo, HI 96720, USA}

\author[0000-0002-0603-3087]{S.~G. Djorgovski}
\affiliation{California Institute of Technology, 1200 E. California Boulevard, Pasadena, CA 91125, USA}

\author[0000-0002-3168-0139]{Matthew J.~Graham}
\affiliation{California Institute of Technology, 1200 E. California Boulevard, Pasadena, CA 91125, USA}

\author[0000-0003-2686-9241]{Daniel Stern}
\affiliation{Jet Propulsion Laboratory, California Institute of Technology, Pasadena, CA 91109, USA}

\author[0000-0001-6746-9936]{Tanya Urrutia} 
\affiliation{Leibniz Institut f\"{u}r Astrophysik, An der Sternwarte 16, D-14482 Potsdam, Germany}

\author[0000-0002-3032-1783]{Mark Lacy}
\affiliation{National Radio Astronomy Observatory, Charlottesville, VA, USA }

\author[0000-0002-7893-1054]{John M.~O'Meara}
\affiliation{Department of Physics, Saint Michael's College, One Winooski Park, Colchester, VT, 05439, USA}

\begin{abstract}
We present the discovery of a gravitationally lensed dust-reddened QSO at $z=2.517$ discovered in a survey for red QSOs by infrared selection. {\em Hubble Space Telescope} imaging in the WFC3/IR F160W and F125W filters reveals a quadruply lensed system in a cusp configuration. We find that compared to the central image of the cusp, the nearby, brightest image is anomalous by a factor of $\sim7-11$. 
Although the source is extremely bright in the mid-infrared, a magnification by a factor of $\sim50-120$ places an upper limit of 1.35 mJy on the intrinsic mid-infrared brightness, well below the $WISE~W4$ detection limit of 6 mJy.
We find that this QSO is moderately reddened, with $E(B-V)=0.7$ and that $\sim1\%$ of the intrinsic spectrum is leaked back into the line of sight resulting in an upturn in its UV spectrum.
We conclude that the QSO's reddening is intrinsic and not due to the lens.   
Consistent with previous red quasar samples, this source exhibits outflows in its spectrum as well as  morphological properties suggestive of it being in a merger-driven transitional phase. 
Depending on how $L_{\rm bol}$ is computed, the quasar's accretion rate may be as high as $0.26~L_{\rm Edd}$.
We detect two Lyman limit systems, at $z=2.102$ and $z=2.431$, with absorption by metal lines likely at small impact parameter to the QSO, and a putative lens redshift of $z=0.599$.
Given the rarity of quad lenses, the discovery of this source allows detailed study of a less luminous, more typical infrared-selected quasar at high redshift.
\end{abstract}


\section{Introduction} \label{sec:intro}

Models of galaxy evolution that invoke major mergers \citep[e.g.,][]{Sanders88,DiMatteo05,Hopkins05} have been highly successful at incorporating the growth of supermassive black holes (SMBH) in galactic nuclei and explaining various scaling relations between the two, such as the $M-\sigma$ relation \citep{Ferrarese00,Gebhardt00}. These models predict a phase during the merger process in which the growing SMBH is enshrouded by dust.  And, while at its peak luminosity, this active galactic nucleus (AGN, or, the more luminous QSO) is heavily obscured and thus elusive to most AGN and QSO surveying techniques, especially those at visible wavelengths. Recent work in the near-infrared has revealed a population of QSOs with moderate amounts of dust extinction that appear to be transitioning from a heavily dust-enshrouded phase, to a typical, unobscured QSO \citep[e.g.,][]{Glikman12,Banerji12,Brusa15}.  

Combining near-infrared and radio has proven to be a very effective method for finding quasars in this transitional state \citep{Glikman04,Glikman07,Urrutia09,Glikman12,Glikman13}. These efforts have resulted in a sample of $\gtrsim120$ dust reddened quasars from the combined FIRST+2MASS surveys (F2M) that spans a broad range of redshifts ($0.1 < z < 3$) and reddenings ($0.1 < E_{B-V} < 1.5$). F2M red quasars were found to be predominantly driven by major mergers \citep{Urrutia08,Glikman15}, are accreting at very high rates \citep{Kim15} and exhibit broad absorption lines associated with outflows and feedback \citep{Urrutia09}. 
Their properties are consistent with buried quasars expelling their dusty shrouds in an an evolutionary phase predicted by merger-driven coevolution models. 

Among the sources in the F2M sample, two gravitationally lensed systems have been identified.  
F2M~J0134$-$0931 is a radio-loud red quasar at $z=2.216$ that is lensed into at least 5 images, possibly by two galaxies at $z=0.7645$ \citep{Gregg02,Winn02,Hall02}. 
This scenario \citep{Keeton03} proposes that the lenses are both spiral galaxies, which may then also be the source of the reddening.  
F2M~J1004+1229, at $z=2.65$, is a rare low-ionization broad absorption line quasar (LoBAL) that includes strong absorption from metastable \ion{Fe}{2} \citep[FeLoBAL;][]{Becker97}.  The location of the reddening in this system is unclear \citep{Lacy02}.

A third high-redshift quadruply lensed quasar, MG J0414+0534 at $z=2.639$, has a heavily reddened continuum as well as red colors among its component images. Several studies have concluded that the majority of the reddening occurs as light passes through a dusty lensing galaxy \citep{Hewitt92,Lawrence95,AngoninWillaime99}.  However, \citet{Curran07} find a low OH to HI column density ratio in the lensing galaxy, suggesting minimal dust and that the reddening is intrinsic to the QSO.

Based on clear color differences between optically-selected lensed quasars and those selected in the radio or infrared, \citet{Malhotra97} suggest that reddening by dust in the lensing galaxy is biasing surveys for lensed quasars, underestimating their numbers.  Alternatively, if the reddening of the quasars is intrinsic, then the population of red quasars may be significantly underestimated. 

Both of the lensed F2M quasars, as well as MG J0414+0534, were selected requiring a radio detection, a wavelength that is largely insensitive to dust reddening, which may have made them easier to find. However, since radio-loud and radio-intermediate quasars make up $\sim10$\% of the overall quasar population \citep{Ivezic02} the radio restriction also limited the sample to a rarer class of quasars. 
In this paper, we report the discovery of a quadruply lensed radio-quiet red quasar discovered in a search for red quasars using {\em WISE} color selection and no radio criterion. 

Throughout this work we quote magnitudes on the AB system, unless explicitly stated otherwise. 
When computing luminosities and any other cosmology-dependent quantities, we use the $\Lambda$CDM concordance cosmology: $H_0=70$ km s$^{-1}$ Mpc$^{-1}$, $\Omega_M=0.30$, and $\Omega_\Lambda=0.70$.

\section{Discovery and Observations} \label{sec:discovery}

\subsection{Selection}\label{sec:selection}

We recently constructed a sample of radio-quiet dust-reddened QSOs selected by their infrared colors in {\em WISE} and 2MASS (W2M), applying well-established color cuts in {\em WISE} color space \citep{Lacy04,Stern05,Donley12,Stern12,Assef13,Assef18,Glikman18} and the infrared-to-optical KX color space \citep{Warren00}.  Our survey covers $\sim 2000$ deg$^2$ with a relatively bright infrared flux limit ($K<16.7$) and has resulted in at least 40 newly-identified red quasars (Glikman et al., in preparation).  Among the sources was W2M~J104222.11+164115.3\footnote{The source name is shortened to W2M~J1042+1641 hereafter.}, whose infrared luminosity, based on {\em WISE} photometry, $L_{IR} = 1.2\times10^{14} L_\odot$, was more luminous than any other known radio-quiet quasar and implied extreme properties suggestive of gravitational lensing. 

The source is undetected in FIRST, implying that its 20 cm flux density is below 1 mJy.  There exists an X-ray source at a separation of 10.4 arcsec in the {\em XMM-Newton} slew survey catalog \citep[XMMSL1;][]{Saxton08} with $0.2-12$ keV flux of $4.0\pm1.8\times 10^{-12}$ erg/s/cm$^2$.  Table \ref{tab:photometry} lists the broadband magnitudes of this source from the surveys used in its discovery.  




\begin{deluxetable}{cc|cc}



\tablewidth{0pt}

\tablecaption{Integrated Photometry of W2M~J1042+1641 from available surveys \label{tab:photometry}}

\tablenum{1}

\tablehead{\colhead{Band} & \colhead{AB Mag} & \colhead{Band} & \colhead{AB Mag}  } 

\startdata
$u$\tablenotemark{a}  &    20.93 $\pm$ 0.10 & $H$\tablenotemark{b}  &    16.87 $\pm$ 0.13 \\
$g$\tablenotemark{a}  &    20.40 $\pm$ 0.03 & $K_s$\tablenotemark{b}  &  15.87 $\pm$ 0.06 \\
$r$\tablenotemark{a}  &    20.26 $\pm$ 0.03 & $W1$\tablenotemark{c}  &   15.52 $\pm$ 0.02 \\
$i$\tablenotemark{a}  &    20.04 $\pm$ 0.04 & $W2$\tablenotemark{c}  &   14.97 $\pm$ 0.02 \\
$z$\tablenotemark{a}  &    18.99 $\pm$ 0.05 & $W3$\tablenotemark{c}  &   13.00 $\pm$ 0.02 \\
$J$\tablenotemark{b}  &    17.84 $\pm$ 0.17 & $W4$\tablenotemark{c}  &   11.84 $\pm$ 0.04 \\
\enddata
\tablenotetext{a}{SDSS Model magnitudes.}
\tablenotetext{b}{2MASS magnitudes.}
\tablenotetext{c}{AllWISE magnitudes.}

\end{deluxetable}

\subsection{Initial Spectroscopy} \label{sec:spec1}

W2M~J1042+1641 was observed with the MODS1B Spectrograph on the Large Binocular Telescope (LBT) observatory for 1200 sec, with the red and blue arms simultaneously, with a 0\farcs6-wide slit on UT 2013 March 14, covering the wavelength range $3300 - 10100\AA$.  After removing the CCD signatures ({\tt modsCCDred}), spectral extraction, wavelength and flux calibration, and telluric correction were done with the IRAF {\tt apall} task. 

On UT 2013 March 19 we observed the source with the SpeX spectrograph \citep{Rayner98} on the NASA Infrared Telescope Facility (IRTF) for 32 minutes using an 0\farcs8-wide slit covering a wavelength range of $0.808 - 2.415$\um.  
Seeing was 1\arcsec\ and sky conditions were clear.  An A0V star was observed within an airmass difference of 0.1 immediately after the object spectrum was obtained to correct for telluric absorption.  The data were reduced using the Spextool software \citep{Cushing04} and the telluric correction was conducted following \citet{Vacca03}.  

\begin{figure*}[ht!]
\plotone{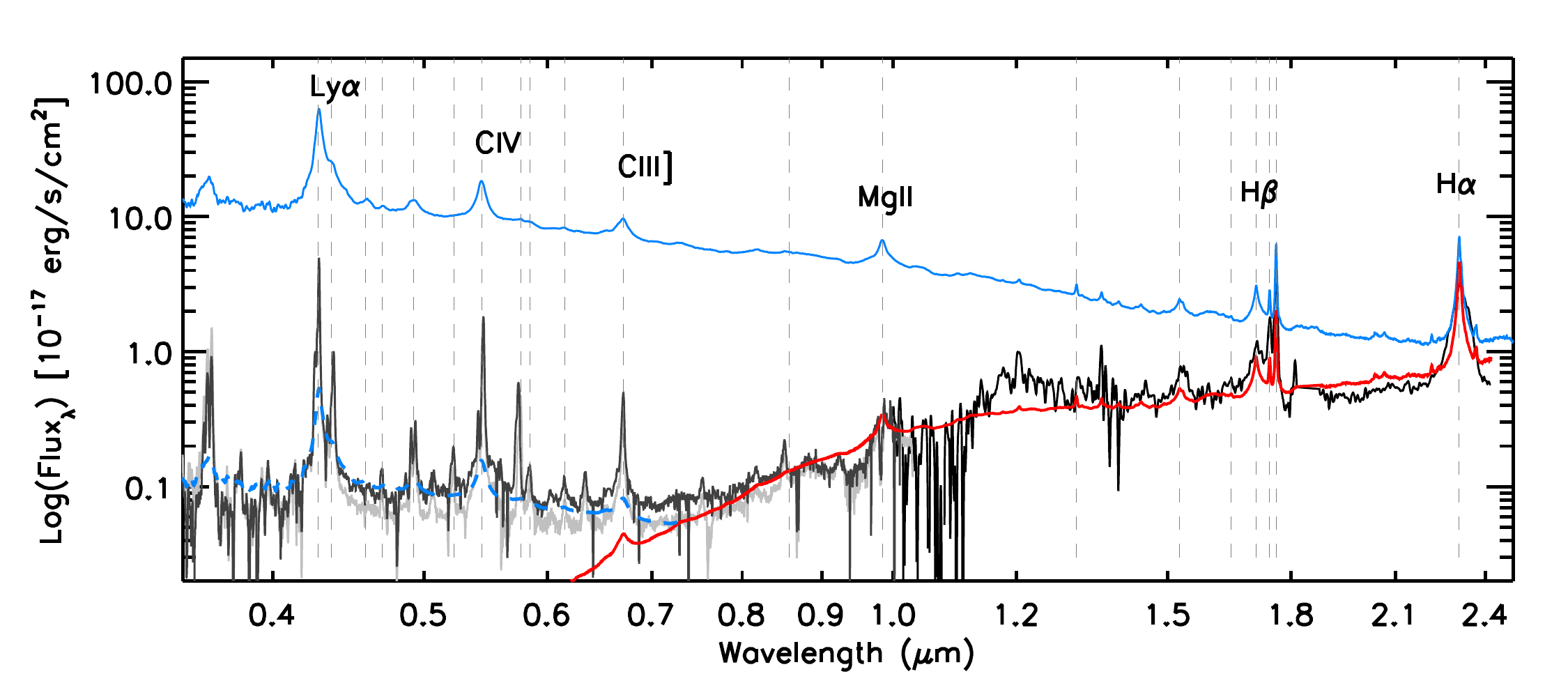}
\caption{Optical-through-near-infrared spectrum of W2M~J1042+1641, plotted on logarithmic wavelength and flux scales. The black line at $\lambda > 1 \mu$m is the near-infrared spectrum, showing broad Balmer line emission shifted to $z=2.517$.  The dark grey line is the LBT optical spectrum and the light grey line is the LRIS spectrum, showing the consistency between the two spectra taken five years apart (observed frame).
Vertical dashed lines mark the locations of strong emission lines seen in the spectrum. 
The red line is the best-fit quasar template (shown in blue) reddened by $E(B-V)=0.68$ to the LBT spectrum combined with the near-infrared spectrum. In both cases, the rest-frame UV part of the spectrum deviates from the reddened template, implying that the obscuring dust is shielding most of the central region but allows $\sim 0.8\%$ of the intrinsic emission to enter out light of sight (blue dashed line), similar to what is observed in Mrk 231 \citep{Veilleux13}. }\label{fig:spec1} 
\end{figure*}

Figure \ref{fig:spec1} shows the combined optical-through-infrared spectrum of W2M~J1042+1641.  The near-infrared spectrum shows strong and broad H$\alpha$ and H$\beta$ plus the narrow [\ion{O}{3}] doublet at $\lambda\lambda4959,5007$, while the optical spectrum shows narrow emission lines in permitted as well as forbidden species, securing a QSO redshift identification of $z=2.517$ 
The blue curve represents an unreddened quasar spectrum, made out of the UV composite quasar template  \citet{Telfer02} combined with the optical-to-near-infrared composite spectrum from \citet{Glikman06}, illustrating the large amount of UV light lost. We fit this curve to the spectrum following the technique outlined in \citet{Glikman07} and find that a suitable fit can only be achieved if the rest-frame UV emission below 2275\AA\ is ignored (observed $\lambda$ 8000\AA).  This best fit is achieved with a quasar template reddened by $E(B-V)=0.68$ (corresponding to a $A_V = 5.4$ mag in the QSO rest frame; red line). 

The excess UV flux, blue-ward of $\sim 8000$\AA, can be explained if the dust were placed close to the AGN, between the broad- and narrow-line-emitting regions. This interpretation is also consistent with a model for the UV spectrum of Mrk 231 suggested by \citet{Veilleux13} in which the broad-line region is reddened by a dusty and patchy outflowing gas. This model predicts a small ``leakage'' fraction of a few percent, which is consistent with a similar degree of leakage seen in the X-ray spectra of other red quasars \citep{Glikman17}.  
A similar conclusion was reached by \citet{Assef16} for a hot Dust Obscured Galaxy (DOG) that displayed blue excess in its spectral energy distribution (SED).  The authors arrive at leaked intrinsic QSO light through a patchy obscuring medium, or by reflection, as the best explanation.

We plot in Figure \ref{fig:spec1} the quasar template scaled to 0.8\% of the intrinsic spectrum (with a dashed blue line) and find that it fits well the spectral shape.  We note that the UV emission lines have a higher equivalent width, but are narrower than the template, similar to `extremely red' quasars in SDSS studied by \citet{Hamann17}.
These arguments lead us to conclude that the dust is local to the lensed quasar and is not reddened by the lens. 

\subsection{Hubble Imaging} \label{sec:hst}

\begin{figure*}[ht!]
\begin{center}
\includegraphics[angle=0,scale=0.75]{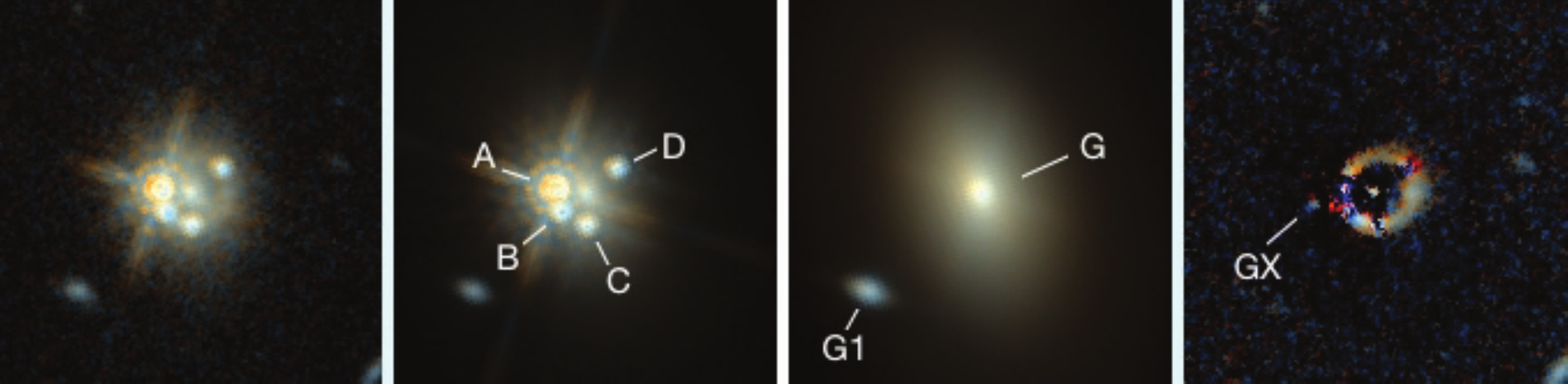}\\
\includegraphics[angle=0,scale=0.75]{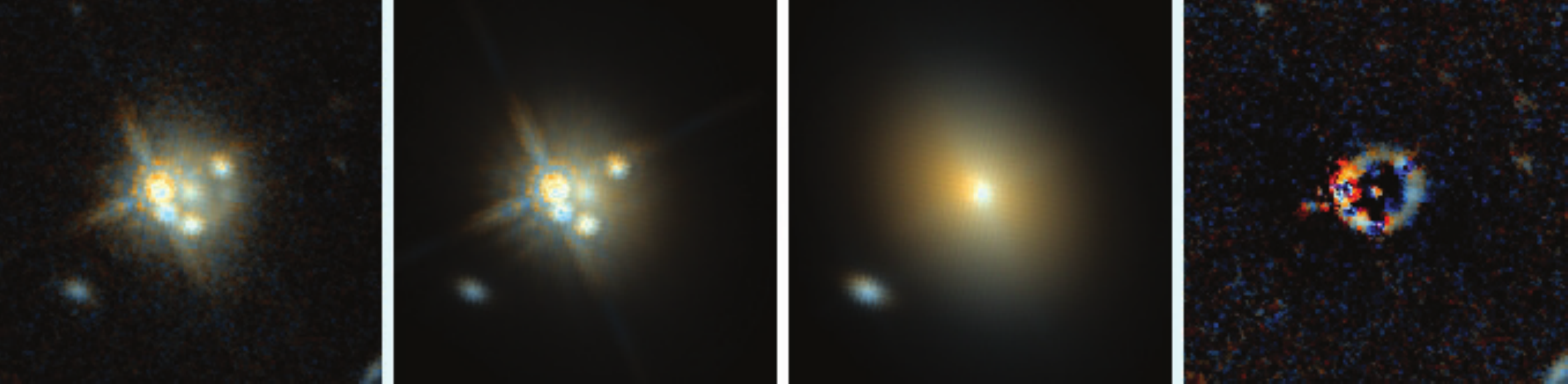}
\caption{Color combined {\em HST} WFC3/IR F125W and F160W images of W2M~J1042+1641 over two visits along with output from a morphological analysis with {\tt galfit} \citep{Peng02}.  The images are oriented with north pointing up and east to the left, the scale is $10\arcsec \times 10\arcsec$, and the visits are rotated by 47.2$^\circ$ with respect to each other. The first column is the drizzled image, $I_{\rm orig}$.  The second column is the best fit model, $I_{\rm mod}$, consisting of four PSFs and a S\'{e}rsic profile.  The third column shows the lensing-galaxy-component of the model in the second column. The fourth column shows the residuals of the model fit, constructed from the difference between the first and second columns, $I_{\rm resid} = I_{\rm orig} - I_{\rm mod}$. The residual shows a clear Einstein ring made up of the lensed quasar's host galaxy light. The modeled components are labeled and referenced in the text.  {\em Top --} Visit 1, UT 2017 February 26. {\em Bottom --} Visit 2, UT May 7.}\label{fig:hst1}
\end{center}
\end{figure*}

We obtained {\em Hubble Space Telescope} imaging of W2M~J1042+1641 with the WFC3/IR camera in Cycle 24 as part of our program to study the host galaxies of W2M red quasars.  We used the F160W and F125W filters, which were chosen to straddle the 4000\AA\ break \citep[following the strategy outlined in ][]{Glikman15}.  We observed the source over two visits, UT 2017 February 26 and UT 2017 May 7, covering both filters in a single orbit observation per visit.  We observed our sources in MULTIACCUM mode using the STEP100 sampling, which is designed to provide a broad dynamic range while avoiding saturation. We performed a 4-point box dither pattern with 400 (224) sec at each position for the F160W (F125W) filter. We reduced the images using the {\tt DrizzlePac} software package to a final pixel scale of 0\farcs06 pixel$^{-1}$.  
The leftmost panel of Figure \ref{fig:hst1} shows the reduced, color-combined images for the two {\em HST} visits ($I_{\rm orig}$), with the first visit shown at the top and the second visit shown on the bottom.  The image reveals four point sources surrounding an extended-appearing source at the center -- this geometry is suggestive of a quadruply lensed quasar. 

We fit $16\arcsec \times 19\arcsec$ cutouts around the system with \texttt{galfit} \citep{Peng02} using a newly constructed point spread function \citep[PSF; following the method described in][]{Glikman15} for each filter. The images from the two visits were taken at different angles, making their combined-image PSF difficult to model. Instead, we fit to the data from each visit separately a model comprised of four PSFs and a S\'{e}rsic profile.
Figure \ref{fig:hst1} shows the results of our profile fits.  
The second column shows the best-fit model ($I_{\rm mod}$) with the four lensed components labeled A, B, C, D, in decreasing order of flux. The third column shows the lensing galaxy model by itself, without the point sources. The fourth column shows the residual image $I_{\rm resid} = I_{\rm orig} - I_{\rm mod}$ where a complete Einstein ring of radius $0.89\arcsec$, originating from the extended lensed quasar host galaxy, is clearly evident. This further verifies W2M~J1042+1641 as a quadruply lensed system.

Finally, Tables \ref{tab:photometry}, \ref{tab:astrometry} and \ref{tab:morphology} list the photometry, relative astrometry of each component, and the morphology of the lensing galaxy, respectively. 




\begin{deluxetable*}{cccccccccc}



\tablewidth{0pt}

\tablecaption{Photometry of the lensing system}

\tablenum{2}

\tablehead{
\colhead{Filter} & \colhead{A (mag)} & \colhead{B (mag)} & \colhead{C (mag)} & \colhead{D (mag)} & \colhead{G (mag)} & \colhead{G1 (mag)} & \colhead{GX (mag)}
} 

\startdata
F125W (1) & 18.18$\pm$0.01 & 20.28$\pm$0.01 & 20.86$\pm$0.01 & 21.45$\pm$0.01 & 19.34$\pm$0.03 & 23.29$\pm$0.05 & $\sim$ 25.7 \\ 
F125W (2) & 18.26$\pm$0.01 & 20.31$\pm$0.01 & 20.69$\pm$0.01 & 21.74$\pm$0.01 & 19.13$\pm$0.04 & 23.34$\pm$0.04 & $\sim$ 25.6  \\ 
F160W (1) & 17.53$\pm$0.01 & 20.05$\pm$0.01 & 20.44$\pm$0.01 & 21.09$\pm$0.01 & 18.73$\pm$0.02 & 23.09$\pm$0.04 & $\sim$ 25.8 & \\ 
F160W (2) & 17.75$\pm$0.01 & 20.09$\pm$0.01 & 20.26$\pm$0.01 & 21.11$\pm$0.01 & 18.74$\pm$0.01 & 23.08$\pm$0.03 & $\sim$ 25.6  \\ 
\enddata


\tablecomments{Photometry has been measured with \texttt{Galfit}. Visits: (1) UT 2017 February 26, (2): UT 2017 May 7. The photometry for image D in visit 2 may be unreliable due to an overlapping diffraction spike from image A.}
\label{tab:photometry} 

\end{deluxetable*}


\begin{deluxetable*}{cccccccccc}



\tablewidth{0pt}

\tablecaption{Relative astrometry of the lensing system}

\tablenum{3}

\tablehead{
\colhead{Axis} & \colhead{A ($\arcsec$)} & \colhead{B ($\arcsec$)} & \colhead{C ($\arcsec$)} & \colhead{D ($\arcsec$)} & \colhead{G ($\arcsec$)} & \colhead{G1 ($\arcsec$)} & \colhead{GX ($\arcsec$)}
} 

\startdata
W $\to$ E & 0.000$\pm$0.000 & 0.152$\pm$0.006 & 0.813$\pm$0.004 & 1.592$\pm$0.005 & 0.792$\pm$0.013 & $-$2.149$\pm$0.034 & $-$0.852$\pm$0.023 \\ 
S $\to$ N & 0.000$\pm$0.000 & $-$0.566$\pm$0.006 & $-$0.909$\pm$0.006 & 0.541$\pm$0.007 & $-$0.076$\pm$0.003 & $-$2.635$\pm$0.003 & $-$0.393$\pm$0.018 \\ 
\enddata


\tablecomments{The error bars represent the scatter in the values measured in the two filters, in both visits.}
\label{tab:astrometry} 

\end{deluxetable*}


\begin{deluxetable}{cccccccccc}



\tablewidth{0pt}

\tablecaption{Morphology of the lensing galaxies}

\tablenum{4}

\tablehead{
\colhead{Object \& Filter} & \colhead{$n$} & \colhead{$R_e$ (\arcsec)} & \colhead{b/a} & \colhead{PA (deg)}
} 

\startdata
G F125W (1) & 5.59$\pm$0.17 & 1.78$\pm$0.11 & 0.60$\pm$0.01 &  14.1$\pm$0.7 \\ 
G F125W (2) &  6.04$\pm$0.19 & 2.20$\pm$0.15 & 0.71$\pm$0.01 &  19.7$\pm$1.0 \\ 
G F160W (1) & 5.47$\pm$0.12 & 1.99$\pm$0.09 & 0.62$\pm$0.01 &  18.0$\pm$0.5 \\ 
G1 F125W (1) & 1.27$\pm$0.35 & 0.21$\pm$0.01 & 0.49$\pm$0.06 &  60.6$\pm$5.1 \\ 
G1 F125W (2) & 1.38$\pm$0.42 & 0.18$\pm$0.01 & 0.51$\pm$0.07 &  70.7$\pm$5.6 \\ 
G1 F160W (1) & 1.40$\pm$0.40 & 0.23$\pm$0.01 & 0.37$\pm$0.05 &  68.9$\pm$3.3 \\ 
G1 F160W (2) & 1.45$\pm$0.32 & 0.22$\pm$0.01 & 0.41$\pm$0.05 &  59.1$\pm$3.1 \\ 
\enddata


\tablecomments{Morphology has been measured with \texttt{Galfit}. Morphology of G in F160W (2) is unreliable due to prominent modeling residuals. Also, the extracted morphology may be affected by the presence of the Einstein ring, which is not included in the model. Angles are positive E of N.}
\label{tab:morphology} 

\end{deluxetable}

\begin{deluxetable}{ccc}



\tablewidth{0pt}

\tablecaption{Lensing mass models}

\tablenum{5}

\tablehead{
 & \colhead{SIE+$\gamma$} & \colhead{SIE+$\gamma$ + GX}
} 
\startdata
$\sigma_G$ & 220.4$\pm$0.6 & 222.3$\pm$0.08 \\
$e=1-b/a$ &  $0.15^{0.03}_{-0.04}$ & $0.32^{0.07}_{-0.08}$ \\ 
$\theta_e$ & $47.3^{6.7}_{-5.4}$ & $26.9^{4.5}_{-3.0}$ \\ 
$\gamma$ & 0.02$\pm$0.01 & 0.04$\pm$0.02 \\ 
$\theta_\gamma$ & $2.6^{17.0}_{-31.3}$ & $-78.7^{7.0}_{-10.5}$  \\ 
$\sigma_{GX}$ & $-$ & $41.5^{6.3}_{-7.4}$ \\ 
$\Delta t$ (days) & 11.7$\pm$1.2 & 18.0$\pm$2.3 \\ 
$\chi^2/\nu$ & 18.4/1 & 0/0 \\ 
\enddata


\tablecomments{Based on fitting with \texttt{glafic}. Uncertainties are determined from 10 Markov chain Monte Carlo (MCMC) runs with 100000-400000 steps each. Both G and GX are assumed to be at $z_l=0.599$, and the source is assumed to be at $z_s=2.517$. Angle are positive E of N. Image D leads, and all other images have similar time delays with respect to it, with differences of less than a day.}
\label{tab:lens} 

\end{deluxetable}


\begin{deluxetable}{ccc}



\tablewidth{0pt}

\tablecaption{Convergence and shear at the location of each image in the SIE$+\gamma$+GX model}

\tablenum{6}

\tablehead{\colhead{Image} & \colhead{$\kappa$} & \colhead{$\gamma$}
} 
\startdata
A & 0.506 & 0.544 \\
B & 0.465 & 0.485 \\
C & 0.611 & 0.539 \\
D & 0.363 & 0.400 \\
\enddata


\tablecomments{}
\label{tab:kappagamma} 

\end{deluxetable}

\subsection{Keck Spectroscopy Along Multiple Position Angles}  \label{sec:lris}

On UT 2018 March 19, we obtained a spectrum with the LRIS spectrograph on the Keck I telescope, orienting the slit along different position angles (PAs) aiming to disentangle the emission from the different components.  We placed a slit along the parallactic angle (79$^\circ$) centered on the brightest component for two 600 s exposures.  Another two 600 s exposures were taken with the slit placed along the A, B, C  components, at a position angle of 41.9$^\circ$.  Finally, a fifth 600 s exposure was performed with a position angle of 128.2$^\circ$ along components B and D including the lensing galaxy with the intention of identifying the redshift of the lens.  Although the seeing was $\sim1\arcsec$, precluding our ability to cleanly separate the different components along the position axis of the slit, the 2D spectrum is clearly extended beyond a PSF. Specifically, the PA $= 128.2^\circ$ data shows two clear lensed AGN components separated by $\sim 1.6$ arcsec. 

The combined LRIS spectrum is shown in light grey in Figure \ref{fig:spec1}.    
The best-fit reddened quasar template to the combined LRIS plus near-infrared spectrum, considering only $\lambda > 8000\AA$, finds $E(B-V)=0.73$ (corresponding to a $A_V = 5.8$ mag in the QSO rest frame).  The LRIS and LBT spectra are remarkably similar, suggesting that not much has changed in this source between the two spectroscopic epochs, 5 years apart in the observed frame, or 1.4 years in the rest frame. This is not unexpected, as significant levels of microlensing and intrinsic variability occur on longer time scales.

\section{Discussion}\label{sec:discussion}

\subsection{Lensing analysis}\label{sec:lensing}

The simplest model known to reproduce the relative positions of the quasar images and lensing galaxy in lensed quasars consists of a singular isothermal ellipsoid with external shear (SIE$+\gamma$). In this model the ``strength'' of the lens is characterized by a velocity dispersion, found to be close to the central velocity dispersion of stars \citep[e.g.,][]{Kochanek94}, which can be estimated for elliptical galaxies from the Faber-Jackson law \citep{faber76}. We fit this model with the code \texttt{glafic} \citep{Oguri10}, using the quasar source redshift of $z_s=2.517$ and assuming a lens redshift of $z_l=0.5985$ (see \S \ref{sec:lensz}). We report the best-fit parameters in Table \ref{tab:lens}. This results in a poor fit with $\chi^2\sim18$ for a single degree of freedom, and suggests that a more flexible model is required. We therefore look at the nearby environment of the system for clues that might explain the poor fit. 

Figure \ref{fig:hst1} shows two nearby structures: G1, located $3.90\arcsec$ from the lensing galaxy, and GX, a structure much fainter but closer to the system ($1.67\arcsec$ from G and $0.94\arcsec$ from A). Including G1 in the fit as a second singular isothermal sphere (SIS) does not result in a significant improvement. In fact, its impact on the model based on its luminosity compared to that of G and scaled by the Faber-Jackson law is negligible, and expected to be even smaller in reality, since it is a spiral\footnote{The S\'{e}rsic index for this galaxy is $n \simeq 1.5$ as noted in Table \ref{tab:morphology}.}. On the other hand, GX is a compact object whose morphology is difficult to constrain, but whose existence as a real object as opposed to a PSF artifact is validated by its presence in both filters and both visits. If we include it in the fit as a SIS at the observed location and at the redshift of the lens (see \S \ref{sec:lensz}), with a velocity dispersion (i.e., lens strength) free to vary, we obtain a perfect fit with zero degrees of freedom. In addition to the quality of the fit, we list the following arguments as to why the SIE$+\gamma$+GX model is more realistic: 

\begin{enumerate}
\item While the velocity dispersion of GX was a free parameter during the fit, its best fit value of $\sim 41.5$ km/s is in excellent agreement with the predicted value of $\sim50$ km/s based on its relative luminosity compared to G and the uncertainty in the Faber-Jackson law.
\item Previous lensing studies find that there is a good alignment between the axes of the light and mass distributions in lensing galaxies, within $\sim 10$ deg \citep[e.g.,][Shajib et al. 2018, in prep.]{Keeton98,Sluse12b}. However, there is less agreement on whether the measured light and mass ellipticities match \citep[e.g.,][]{Sluse12b,Gavazzi12} or not \citep[e.g.,][Shajib et al. 2018, in prep.]{Keeton98,Rusu16}. We find that when GX is added to the model, the mass and light profiles of the main lens G match in terms of both alignment and ellipticity. The matching of the ellipticity is likely a consequence of the Einstein radius for this system being well inside the effective radius of the lensing galaxy. This is because the light profile is dominated by baryons, and gravitational lensing is most sensitive to the mass profile at the Einstein radius, thus also in the baryon-dominated region.  
\item Like any cusp configuration in a smooth lensing mass model, the central image (in this case B) is expected to have a flux equal to the sum of the two surrounding images \citep[A and C; e.g.,][]{Keeton03}. This makes the observed flux of image A anomalous by more than an order of magnitude compared to the SIE$+\gamma$ model. Depending on which filter/visit is used to measure the observed flux, this anomaly exceeds the largest anomaly recorded in the optical, for SDSS~J0924+0219 \citep{Inada03}. The addition of GX boosts the flux of image A in relation to B by a factor of $\sim1.5$, thus mitigating some of the discrepancy (see Figure \ref{fig:lens}).
\end{enumerate}

\begin{figure}[ht!]
\epsscale{1.3}
\plotone{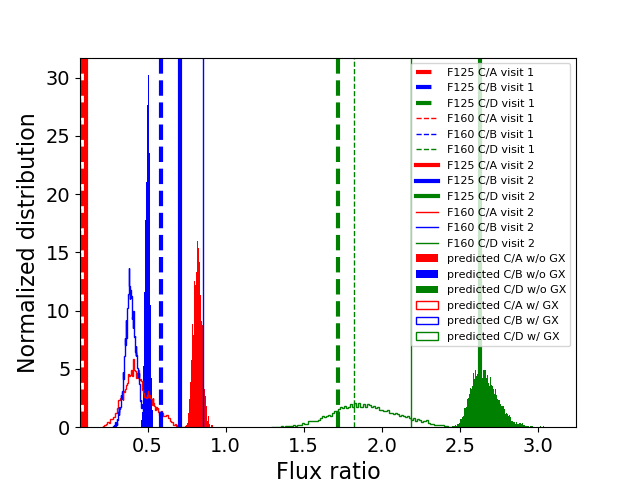}
\caption{Comparison of observed and predicted flux ratios. The histograms consist of 10000 points from the MCMC chains computed with \texttt{glafic}, and the vertical lines correspond to the photometry measured by \texttt{Galfit}. The observed flux ratios C/A (much smaller than the model prediction) and C/B (larger than predicted) are anomalous. Visits: (1) UT 2017 February 26, (2): UT 2017 May 7. The fluxes of image D in visit 2 may be unreliable due to a diffraction spike from image A falling on top of it.
}\label{fig:fluxratio} 
\end{figure}

\begin{figure}[ht!]
\epsscale{1.4}
\plotone{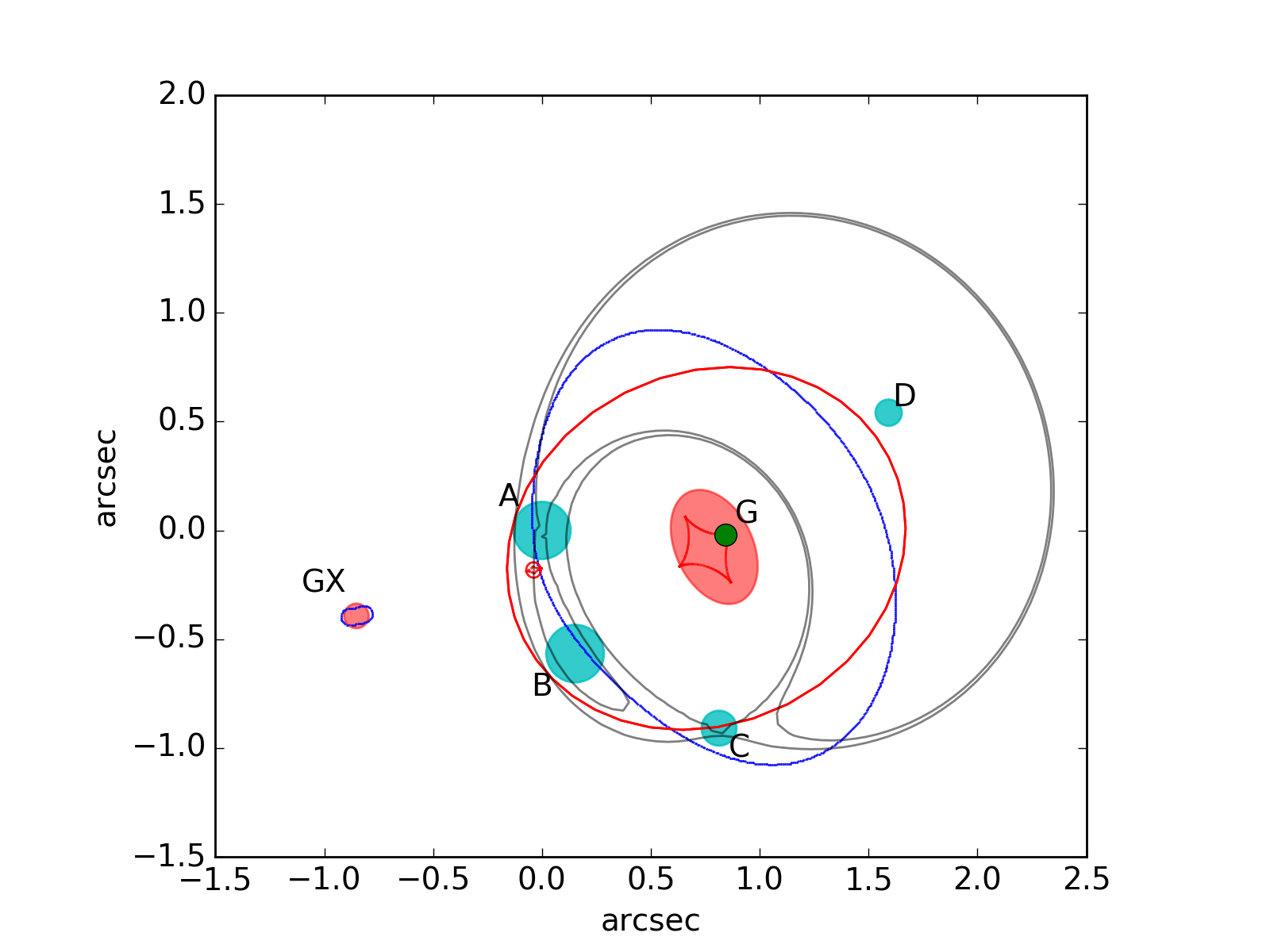}
\caption{Lensing configuration for the SIE$+\gamma$+GX model. The two lenses are marked with ellipses whose sizes scale in proportion to the modeled velocity dispersions. The ellipsis for G shows the orientation and axis ratio predicted by the mass model. Critical lines are drawn in blue and caustics in red. The green circle marks the location of the source quasar in the source plane, and the cyan circles mark the locations of the observed images. The size of the circles is in proportion to the predicted flux ratios. Surface isochrones at the location of the saddle point images A and D are drawn in gray. 
}\label{fig:lens} 
\end{figure}

An important quantity to derive for the purpose of characterizing the physical properties of the source is its total magnification. The SIE$+\gamma$+GX model predicts a total magnification (i.e., integrated over the flux of the four images) of $53.3\pm5.3$. However, this can be significantly affected by flux anomalies, especially in the case of image A, which is much brighter than predicted. We plot in Figure \ref{fig:fluxratio} the observed (vertical lines) versus model-predicted (histogram) flux ratios, with image C relative to the other images. It is apparent from this figure that both C/A (plotted in red) and C/B (blue) are anomalous, whereas C/D (green) matches well the SIE$+\gamma$+GX model prediction. We therefore use the flux of image C, the brightest among C and D, as an anchor to estimate the total magnification, as $\mu_\mathrm{total}=obs_\mathrm{A/C}\times\mu_\mathrm{C}+obs_\mathrm{B/C}\times\mu_\mathrm{C}+\mu_\mathrm{C}+\mu_\mathrm{D}$. Here $obs$ refers to the observed flux ratio and $\mu$ is the model-predicted magnification for each individual image. Propagating the observed and model uncertainties, we obtain $\mu_\mathrm{total}=122\pm26$.

Depending on the filter and visit where A/B is measured, this flux ratio is anomalous by a factor of 7-11. From Figure \ref{fig:fluxratio} we see that C/A is also anomalous, and therefore reported to the reference image C the flux anomaly factors are 4-6 for A/C and 1.5-2.2 for C/B. We note that this is contrary to the case of SDSS~J0924+0219, where the anomalous flux due to a combination of substructure and microlensing \citep[][and references within]{Macleod15} is suppressed rather then boosted, by a factor of $\sim12$ \citep{Keeton06}. The fact that A, the most flux anomalous image, is a saddle point of the time arrival surface is consistent with the findings of \citet{Schechter02}, that saddle points are more susceptible to microlensing effects. Still, there is no compelling evidence to definitively attribute the anomalous flux ratios to microlensing. In fact, in \S \ref{sec:mbh} we show that our estimates of the black hole mass from the integrated fluxes in F160W and {\it WISE} W4 show better agreement if the larger of the two magnification factors computed above is used for the measurement from W4. This would suggest that the large magnification factor persists at long wavelengths, and is therefore due to substructure rather than microlensing. However, uncertainties in the assumed reddening  and bolometric corrections make this result less reliable (see \S \ref{sec:mbh}). We also note that we do not have reliable spatially resolved spectroscopy to look for chromatic microlensing variations \citep[e.g.,][]{Sluse12a}, and exploring color differences between the quasar images in the two imaging filters is tempered by variations between the fluxes in the two visits. We attribute these variations to modeling difficulties (e.g., the diffraction spike on top of image D in visit 2; also the contribution from the light in the Einstein ring) and to intrinsic variability. 

To facilitate the theoretical investigation of microlensing lightcurves in this system we report the values of convergence and shear at the location of each image in Table \ref{tab:kappagamma}. Finally, while the integrated spectrum does show an enhancement in the flux in the blue, we have argued in \S \ref{sec:spec1} that this can be explained by factors intrinsic to the quasar, and does not require microlensing.
 
\subsection{Structure along the Einstein ring}\label{sec:host}

\begin{figure}
\begin{center}
\includegraphics[angle=0,scale=0.7]{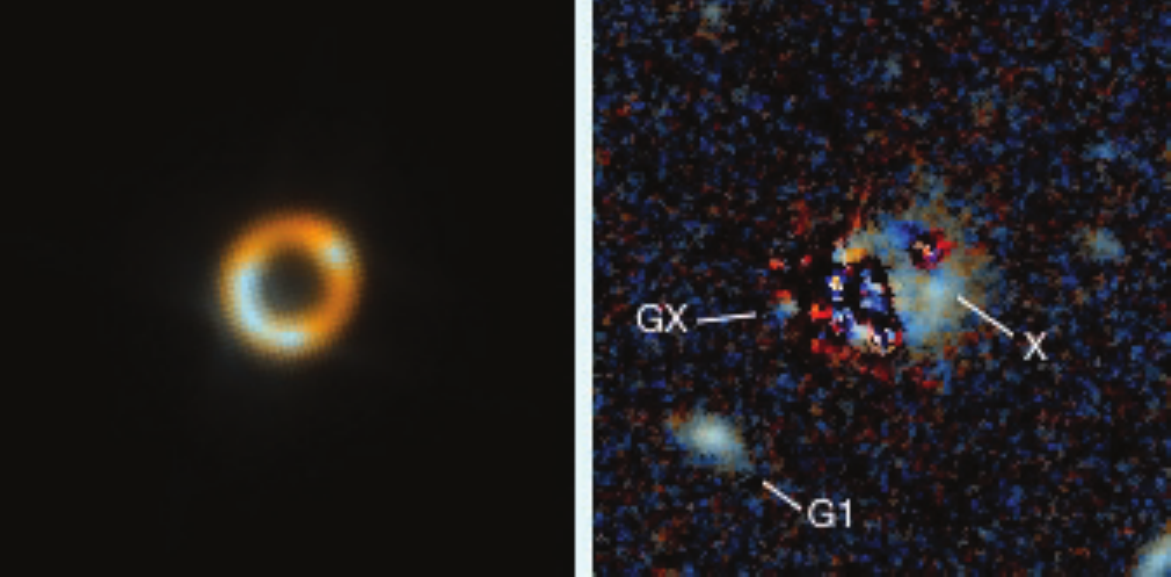}\\
\includegraphics[angle=0,scale=0.7]{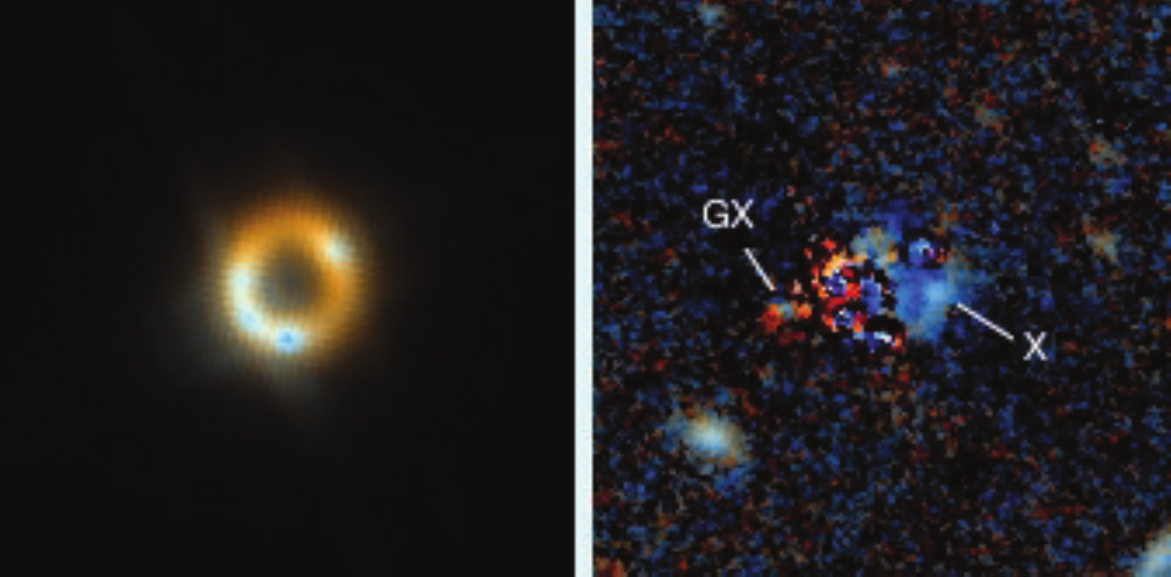}
\caption{Color combined images of the modeled (left column) and subtracted (right column) Einstein ring for  visit 1 (top row) and visit 2 (bottom row). The modeling is performed with \texttt{hostlens} for the best-fit SIE+$\gamma$+GX mass model, and uses an extended source modeled as a circular S\'{e}rsic profile. Same orientation and scale are used as in Figure \ref{fig:hst1}.}\label{fig:hst2}
\end{center}
\end{figure}

The Einstein ring seen as residual in Figure \ref{fig:hst1} does not appear to be smooth, but instead shows at least one blueish peak present in both visits, South of image D. We used \texttt{hostlens} \citep{Rusu16} to model the smooth component of the Einstein ring, and we show the resulting fit and residuals in Figure \ref{fig:hst2}. Here, the blueish peak is marked with X. 

Ignoring the remote possibility that it is a star in the Galaxy, this object has two possible interpretations: First, similar to GX, it could be a satellite of the main lens galaxy. We note that the lensing model can still reproduce exactly the positions of the quasar images even if X is included in the mass model as a second perturber with velocity dispersion estimated from the Faber-Jackson law. However, this interpretation is unlikely for two reasons: i) it is located at about half an effective radius from G, a region where substructure is unlikely to survive; ii) its location on top of the Einstein ring argues towards it being associated with the source. In this second interpretation, X would be a close companion of the quasar host galaxy. Lens modeling implies a distance from the host of $\sim 1.1$ kpc, but predicts three more images, the brightest pair being located very close to image C. The close proximity may make these difficult to separate from the residuals of the fit. Object X is not the only structure visible in the residuals, which show additional blue diffuse light. Given the small separation from the host, these structures may correspond to patches of star formation in the otherwise red quasar host (as seen from the color of the overall Einstein ring). This may suggest a recent merger, which would be in agreement with most of the red quasars being mergers \citep{Urrutia08,Glikman15}.

A future paper will further explore these possibilities by conducting a more detailed analysis of the Einstein ring, contrasting different techniques of modeling the extended source emission (Rusu et al., in preparation). Finally, we note that simultaneously fitting the Einstein ring and the point-like quasar images does affect the measured flux ratios reported in Table \ref{tab:photometry} and Figure \ref{fig:fluxratio}, but not enough to affect the interpretation in \S \ref{sec:lensing}.

\subsection{Spectral characteristics}\label{sec:qso}

\begin{figure*}[ht!]
\plotone{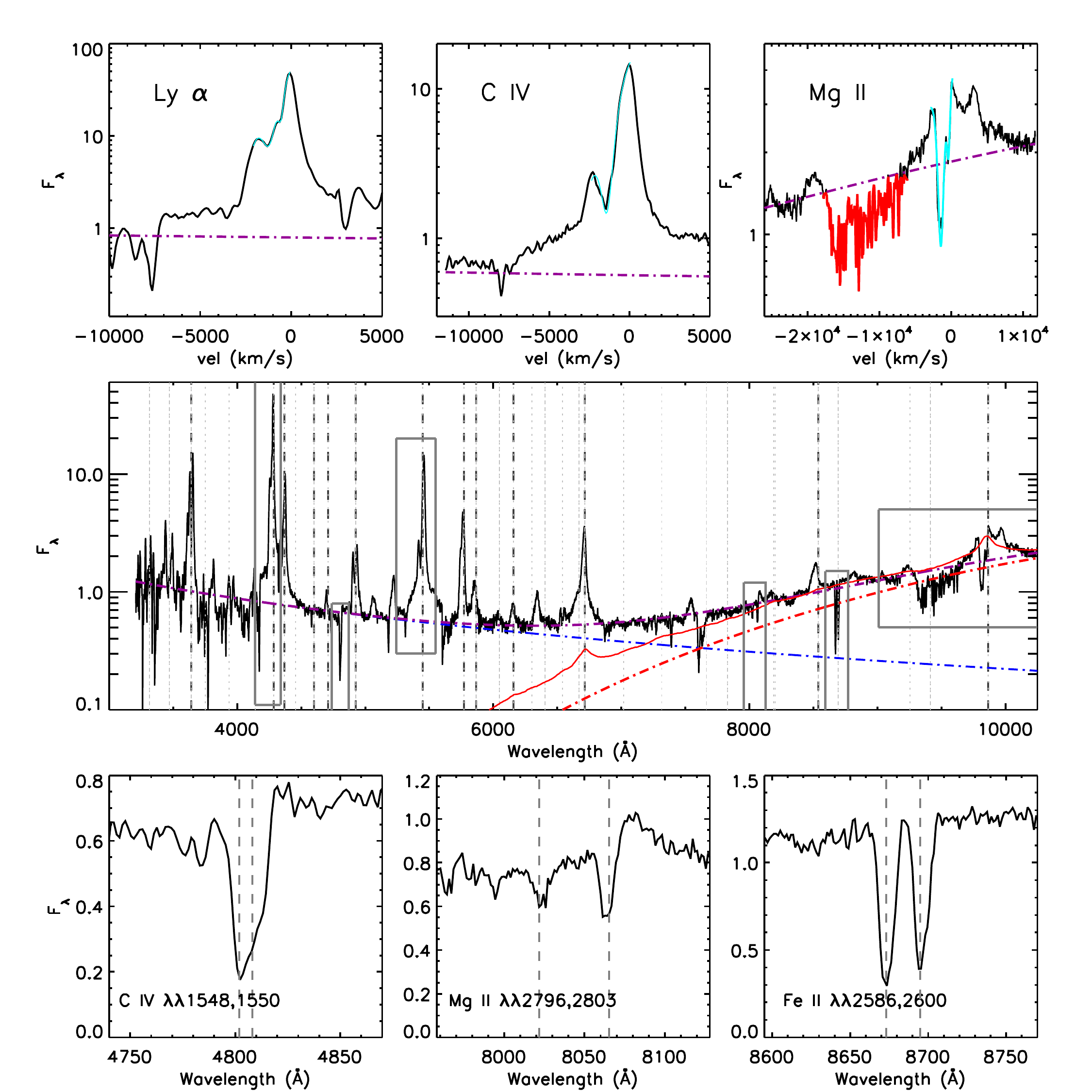}
\caption{Deep LRIS spectrum of W2M~J1042+1641 (middle panel; black line) with typical (light grey) and prominent (dark grey) QSO emission lines marked with vertical dotted and dashed lines, respectively.  A best-fit reddened power law continuum plus an unreddened leakage component is shown with a purple dash-dot line, with the reddened ($E(B-V)=0.89$) and leaked components plotted in red and blue, respectively.  The solid red line is the best-fit reddened template with $E(B-V)=0.73$.  
{\em Top row --} Profiles of prominent emission lines \ion{Mg}{2}, \ion{C}{4}, and Ly$\alpha$ (left to right) plotted in velocity space.  All three lines show a distinct absorption feature, that is well fit by a single or double Gaussian profile, with an outflow speed of $\sim 1000 - 1500$ km s$^{-1}$.
We indicate the presence of a BAL feature in \ion{Mg}{2} emission line (red) that is absent in the other lines.
{\em Bottom panel --} Absorption line profiles of a Lyman-limit system at $z=2.1$.
The line regions plotted in the top and bottom columns are also marked with grey boxes in the middle panel.
}\label{fig:LRIS} 
\end{figure*}

The LRIS spectrum of W2M~J1042+1641, taken across multiple PAs, for a total of 50 minutes (5 exposures $\times 10$ minutes each) is our best dataset for investigating the QSOs emission line properties and searching for intervening absorption systems, including from the lensing galaxy.  Figure \ref{fig:LRIS} shows the combined LRIS spectrum of W2M~J1042+1641 (black line) plotted on a logarithmic y-axis in order to enhance features across its dynamic range.  

To identify and study line features, we require a better determination of the QSO continuum.  Following our interpretation that W2M~J1042+1641 is a reddened QSO with some leakage of the intrinsic spectrum, we model the QSO spectrum as a power law, with spectral index $\alpha_\nu = -0.5$ ($F_{\lambda0} \propto \lambda^{-(\alpha_\nu+2)}$). We then fit the line-free portions of the spectrum with a two-component power-law, one reddened and one pure, both with the same power-law index:
\begin{equation}
F_\lambda = A F_{\lambda 0}  + B F_{\lambda 0} e^{-\tau_\lambda}. \label{eqn:cont}
\end{equation}
The best fit is shown with a purple dash-dot line along with the unreddened component, plotted with a blue dash-dot line, and the reddened component, plotted with a red dash-dot line.  For comparison, we also show the best-fit reddened quasar template (red solid line) that was determined from the near-infrared spectrum combined with the LRIS spectrum, using only wavelengths longer than 8000\AA.  While the template-based fit results in $E(B-V)=0.73$ mag, the two-component power-law fit yields $E(B-V) = 0.89$ mag.  

The dilution of the continuum in the rest-frame UV allows us to see many emission lines not typically seen in luminous QSO spectra at unusually high equivalent widths\footnote{This is also apparent in Figure \ref{fig:spec1}, where the spectral lines appear stronger than the composite QSO template that has been scaled to match the reddened QSO continuum.}. We mark with grey vertical dotted lines the positions of all emission lines reported in the SDSS quasar composite spectrum from \citet{VandenBerk01}.  Thick dark grey lines highlight prominent emission features in this spectrum and include Ly$\beta$~$\lambda$1033, Ly$\alpha$~$\lambda$1216, \ion{N}{5}~$\lambda$1240, \ion{O}{1}~$\lambda$1305, \ion{C}{2}~$\lambda$1336,\ion{Si}{4} + \ion{O}{4} ~$\lambda$1398,  \ion{C}{4}~$\lambda$1546, \ion{He}{2}~$\lambda$1638, \ion{O}{3}]~$\lambda$1665, \ion{N}{3}]~$\lambda$1748, \ion{C}{3}]~$\lambda$1906, [\ion{Ne}{4}]~$\lambda$2423,  \ion{Mg}{2}~$\lambda$2800.

We also see lines at 5062\AA, 5221\AA,  6340\AA, and 7540\AA \ that are not listed in \citet{VandenBerk01} as common QSO lines. If the lines are emitted at the QSO redshift, then we might be seeing \ion{N}{4}]~$\lambda$1486 ($\Delta\lambda \simeq -4\AA$) which is a rare QSO line seen in nitrogen enhanced quasars \citep{Bentz04}, and low luminosity, high-redshift quasars \citep{Glikman07b}. \ion{N}{4}] is also associated with low-metallicity starbursts \citep{Fosbury03}, and may come from the star forming knots in the host galaxy (possibly seen in the Einstein ring \S \ref{sec:host}). 

We see absorption in the complex-appearing \ion{Mg}{2} line, plotted in the right-hand panel of Figure \ref{fig:LRIS} in velocity space.  The purple line is the continuum from our fit.  We see broad absorption with blueshifted velocities between $-6000$ and $-18000$ km s$^{-1}$ (red) which likely consists of many outflowing systems, as the feature has no clean profile.  We also see two other distinct absorption systems that are well fit by a double Gaussian, with Full Width at Half Maximum (FWHM) speeds of 4348 km s$^{-1}$ and 1838 km s$^{-1}$, and systemic outflow speeds of $-1407$ km s$^{-1}$ and $-350$ km s$^{-1}$.
A feature peaking at +3000 km s$^{-1}$ from the \ion{Mg}{2} position is unidentified and may be part of  the \ion{Mg}{2} line itself, whose profile is complex.

The next two panels of Figure \ref{fig:LRIS} display the same for \ion{C}{4} (middle) and Ly$\alpha$ (left), which both show a blueshifted absorption feature, also well fit by a Gaussian.  The feature at \ion{C}{4} is best fit by a single Gaussian with FWHM velocity width of 3060 km s$^{-1}$ and an outflow velocity of $-1576$ km s$^{-1}$. The absorption at Ly$\alpha$ is best fit by two Gaussians with FWHM velocity widths of 4184 km s$^{-1}$ 1306 km s$^{-1}$, outflowing at $-1013$ km s$^{1}$ and $-515$ km s$^{-1}$. 

These features are indicative of outflowing gas, likely related to each other, given their similar characteristics in each line. Interestingly, the BAL feature is only seen near \ion{Mg}{2} but appears to be absent near \ion{C}{4} and Ly$\alpha$.  This is unusual for LoBAL spectra, which typically have narrower and shallower \ion{Mg}{2} troughs as compared with \ion{C}{4} \citep{Trump06}.  One explanation for this observation is that the absorber is affecting the reddened component and is thus too weak to be seen in \ion{C}{4}.
Red quasars have an anomalously high fraction of BALs, particularly LoBALs \citep[37\% compared with $\sim 1\%$][]{Urrutia09,Trump06}.  It is possible that the feature is present, but has a weak equivalent width due to the weak continuum at \ion{C}{4} and Ly$\alpha$, and is further diluted by the superposition of the leaked component.

We also identified multiple intervening absorption systems including Lyman-limit absorption at $z=2.431$ identified in \ion{H}{1} $\lambda$1216, \ion{C}{2} $\lambda$1335, \ion{C}{4} $\lambda\lambda$1548,1550, and a second Lyman limit system at $z=2.102$ through deep absorption in the \ion{Fe}{2} $\lambda\lambda$2586,2600, \ion{Mg}{2} $\lambda\lambda$2796,2803, and \ion{C}{4} transitions. 
The absorption features from the second system are shown in the bottom panel of Figure \ref{fig:LRIS}
We also identify a tentative absorber at $z=1.179$ via \ion{Mg}{2}.
 
\subsubsection{Lens Redshift} \label{sec:lensz}

The LRIS frame oriented along the host galaxy (PA$ = 128.2^\circ$) clearly shows two AGN components separated by a few pixels. We detect no clear signal from the lens itself.  

\begin{figure}[ht!]
\plotone{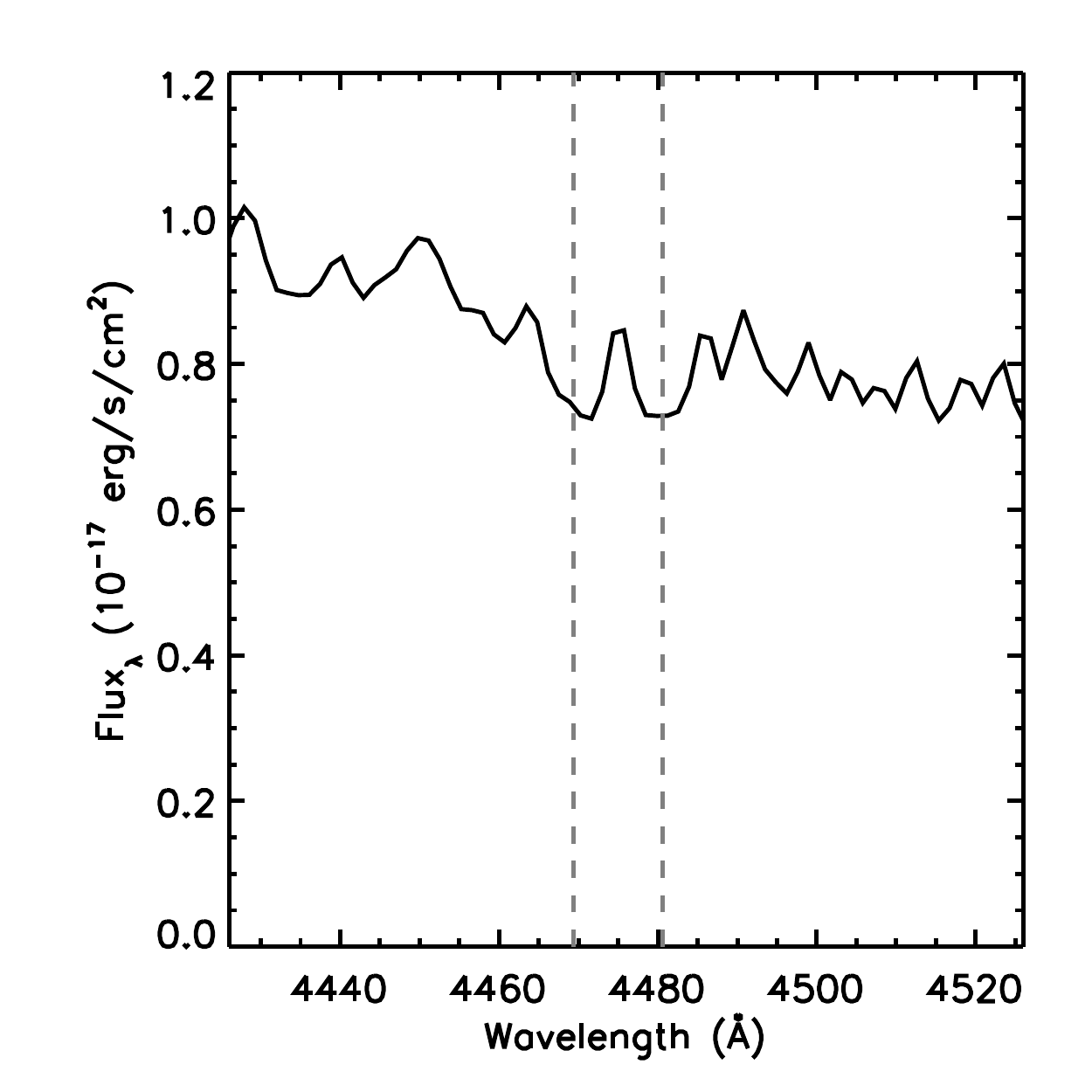}
\caption{We identify a \ion{Mg}{2} $\lambda\lambda$2796,2803 absorption feature, marked by dashed vertical grey lines, at a redshift of $z=0.5985$.}\label{fig:mg2}
\end{figure}

We identify absorption consistent with \ion{Mg}{2} $\lambda\lambda$2796,2803 at $z=0.5985$, Figure \ref{fig:mg2}.  This absorption may be contaminated by \ion{O}{1} $\lambda$1302 / \ion{Si}{2} $\lambda$1304 absorption from the $z=2.431$ Lyman-limit system, but is in food agreement with the photometric redshift we estimate for the lensing galaxy using an early-type template, and we thus adopt this redshift as the redshift for the lensing galaxy.

\subsection{Black hole mass} \label{sec:mbh}

The H$\alpha$ line seen in the near-infrared spectrum is well fit by a single Gaussian, with a FWHM in velocity space of $v_{\rm FWHM} = 8564$ km s$^{-1}$, which we use to estimate the black hole mass of W2M J1042+1641.  We combine the line width with an estimate of the QSO's intrinsic luminosity at 5100\AA\  and apply them to the single-epoch virial black hole mass estimator ($M_{\rm BH,vir}$) following the formalism of \citet{Shen12},

\begin{equation}
\log \bigg(\frac{M_{\rm BH,vir}}{M_\odot} \bigg) = a + b \log \bigg(\frac{L_{5100}}{10^{44} \rm erg/s} \bigg) + c \log \bigg(\frac{v_{\rm FWHM}}{\rm km/s} \bigg),
\end{equation}
with the values $a=0.774$, $b=0.520$, $c=2.06$  derived by \citet{Assef11}. 
We choose the H$\alpha$ line, because it is in a region of minimal reddening in our spectrum and the more-commonly-used H$\beta$ is strongly blended with [\ion{O}{3}].  

To determine $L_{5100}$, we turn to the best-fit continuum that we derived in \S \ref{sec:qso} that traces well the continuum along the F160W filter in the rest-frame out to $\lambda_{\rm rest} = 5100$\AA. In addition, because the magnification of the QSO is derived from the F160W image, de-magnifying the observed luminosity at this wavelength ($\lambda_{\rm F160W, rest} = 4370$\AA) will give the most internally consistent results.  We estimate the continuum flux, represented by Equation \ref{eqn:cont}, through the F160W band, and scale it to the summed flux from the four QSO components listed in Table 2.  
We find the intrinsic flux at this wavelength by de-reddening the second term in Equation \ref{eqn:cont} and adding to it the first term.  We scale the observed flux from the four QSO components to match the de-reddened continuum and then shift their flux value at $\lambda_{\rm F160W, rest} = 4370$\AA\ to the de-reddened flux at $\lambda_{\rm rest} = 5100$\AA.  Finally, we divide the flux it by a magnification factor of 122 (\S \ref{sec:lensing}) and compute a luminosity of $\log(L_{5100} ~{\rm [erg/s]) = 41.65}$.  These values yield $\log (M_{\rm BH,vir} ~[M_\odot]) = 7.54$.  

To estimate $L/L_{Edd}$, we use a bolometric correction (BC) at 5100\AA\ of 9.5 from \citet{Richards06}. This gives $\log(L_{\rm bol} ~ {\rm [erg/s]}) = 42.63$ and a corresponding Eddington ratio, $L/L_{Edd} = 0.001$, which is a comparatively low accretion rate compared with more luminous red quasars, most of which have $L/L_{Edd} \gtrsim 0.5$ \citep{Kim15}.

We now recompute these physical parameters, but by using the observed {\em WISE} $W4$ rather than the F160W flux. On the one hand, the calculation for $L_{5100}$ using the {\em WISE} $W4$ flux requires no need for a reddening correction\footnote{An extinction of $E(B-V) = 0.7$ results in $A_{W4}=0.06$ mag (or 6\% of the flux) at the rest wavelength of 6.28\um.}. On the other, the magnification factor relevant at this wavelength would be the one observed in the {\it Hubble} images ($\mu\sim122$) if the flux anomaly reported in \S \ref{sec:lensing} is due to substructure, but would be much closer to the model-predicted value of $\mu\sim53$ if it is instead due to microlensing, which is expected to become insignificant at long wavelengths. We scale the observed {\em WISE} $W4$ flux to the flux at 5100\AA\ using the SED from \citet{Richards06} for optically red quasars. We de-magnify this flux by a factor of 53 and find $\log (L_{5100} ~{\rm [erg/s]}) = 42.57$, resulting in $\log (M_{\rm BH,vir} ~[M_\odot]) = 8.02$.

Using a BC of 7.5 from \citet{Richards06} based on the optically red quasar SED\footnote{We note that using the template for all SDSS quasars (BC$\simeq7.9$) does not significantly affect the results.} for a frequency corresponding to $W4$ (22 $\mu$m) in the rest frame, $\log (\nu {\rm [Hz]}) = 13.68$, we compute $\log (L_{\rm bol} ~{\rm [erg/s]}) = 45.47$ -- significantly higher than the estimate based on the de-reddened and F160W flux (still, using $\mu = 53$). We find an Eddington ratio of $L/L_{Edd} = 0.225$.  
If we use a higher magnification at $W4$ ($\mu = 100$) the estimated black hole mass reduces to $\log (M_{\rm BH,vir} ~[M_\odot]) = 7.66$ -- more consistent with the value derived from the {\em HST} photometry.  The bolometric luminosity is also reduced, to $\log (L_{\rm bol} ~{\rm [erg/s]}) = 45.17$, yielding $L/L_{Edd} = 0.26$.  

The $L/L_{Edd} \sim 0.25$ derived from $W4$ is more consistent with the values found for other red quasars, and may point to enhanced infrared emission in these sources, even compared to the red quasars used to construct the SEDs in \citet{Richards06}. We consider these two approaches to computing $M_{\rm BH}$ and $L/L_{Edd}$ to span the possible range for this complex system.  The largest uncertainty in the first method (using F160W) is the corrected 5100\AA\ flux that depends strongly on the reddening estimate for the continuum. The largest uncertainty in the second method is the magnification factor, $\mu$.

\subsection{Magnification and population analysis}\label{sec:population}

\begin{figure}[ht!]
\plotone{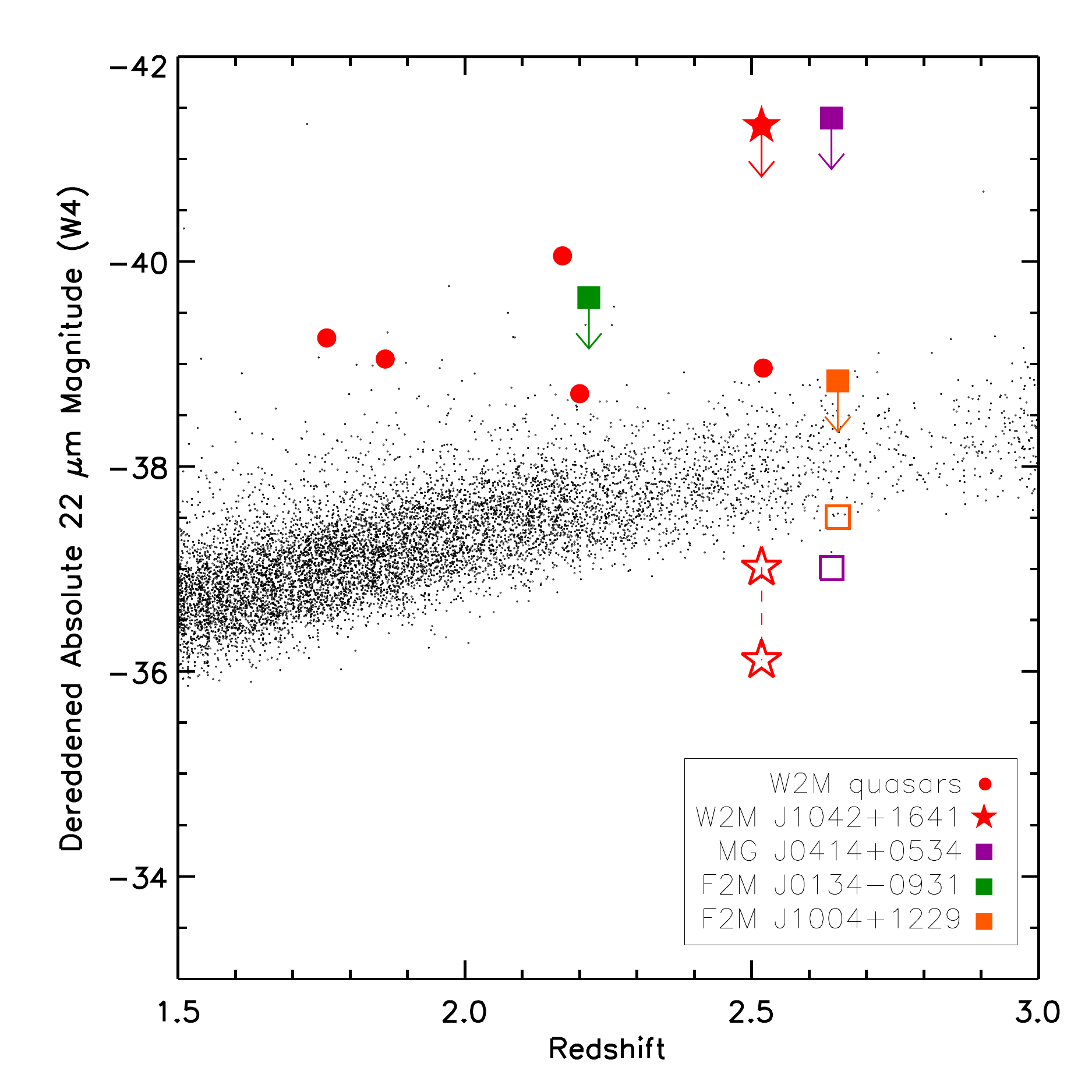}
\caption{Absolute magnitude at 22\um, from {\em WISE}, for the sample of red quasars in the W2M survey.  The observed luminosity of W2M~J1042+1641 is indicated by a star symbol, with a downward-facing arrow pointing toward its intrinsic luminosity whose range is shown by two open red star symbols connected by a dashed red line.  This figure indicates that W2M~J1042+1641 is a reddened quasar drawn from a population of sources that is not accessible to the wide-area quasar sample currently being studied. We show, for comparison, the observed and magnification-corrected luminosities of the other three known high-redshift lensed red quasars discussed in the text, using only an upper limit for F2M~J0134-0931, whose lensing configuration is too complex to find a reliable magnification \cite{Keeton03}.}
\label{fig:lumz}
\end{figure}

Figure \ref{fig:lumz} shows W2M~J1042+1641 (red filled star) on a WISE $W4$ luminosity versus redshift diagram. The filled red circles are the other high-redshift W2M quasars found in our study, which is spectroscopically complete (Glikman et al., in preparation). We compare them with $\sim 20,000$ quasars from SDSS with matches in the UKIDSS and WISE catalogs with spectroscopic redshifts from \citet{Peth11} (black dots). 
The de-magnified position of  W2M~J1042+1641 is shown as an open star, below the lower envelope of SDSS quasars detected which represents the WISE $W4$ detection limit. We see that this red quasar exists among a population whose luminosity is too low to be studied in wide-field surveys that yield the vast majority of quasars in the literature.  

We also plot in Figure \ref{fig:lumz} the positions of the three other lensed red quasars: MG J0414+0534 \citep[purple square;][]{Lawrence95}, F2M J0134$-$0931 \citep[green square;][]{Gregg02}, and F2M J1004+1229 \citep[orange square;][]{Lacy02}.  
All three of these quasars have {\em HST} images in the archives, and were analyzed by fitting SIE$+\gamma$ models to the relative astrometry reported in CASTLES\footnote{CfA-Arizona Space Telescope LEns Survey, C.S. Kochanek, E.E. Falco, C. Impey, J. Lehar, B. McLeod, H.-W. Rix, \url{https://www.cfa.harvard.edu/castles/}}, in order to determine their magnification factors.
Their de-magnified positions are shown with corresponding open symbols and they, too, lie at the edge of or below the faintest luminosities accessible to SDSS, UKIDSS, and WISE. 

The W2M survey finds 7 high redshift ($z>1.7$) QSOs, including the lensed quasar F2M~J1004+1229, which we recover from our previous FIRST-selected sample. That means that the lensing fraction of this complete, flux-limited sample is 2/7 = 28\% -- three orders of magnitude higher than the lensing fraction for luminous unobscured quasars in typical surveys \citep{OM10}. We note that a similar fraction (24\%) is found by \citet{Donevski18} among Herschel-selected `500 \um-risers' -- a rare population of star forming galaxies, and is explained by very steep counts and large source redshifts, boosting the lensing optical depth \citep{Negrello07}. The lensing fraction of the F2M survey (Section \ref{sec:intro}) is also large, at $\sim2$\%, but smaller than that of the W2M survey. This is likely due to the F2M survey being about a magnitude deeper than W2M.

The de-magnifed luminosity densities computed from the $W4$ flux density of W2M~J1042+1641 ($\lambda_{\rm rest} = 6.2$ \um) are $L_{6\mu{\rm m}} = 6.8\times10^{32}$ erg s$^{-1}$ Hz$^{-1}$ for magnification = 53.  However, when considering a magnification of 122, we compute $L_{6\mu{\rm m}} = 2.8\times10^{32}$ erg s$^{-1}$ Hz$^{-1}$.
This luminosity is near the knee of the red quasar sample found in deep {\em Spitzer} fields (Fig.~1; \citealt{Lacy15} and Fig.~19; \citealt{Glikman18}) and is thus on the bright-end side of the red quasar luminosity function (QLF). Although \citet{Glikman18} found that the bright-end slopes of the QLF for red and blue quasars are very similar ($\gamma_{2, {\rm red}} = 2.44\pm0.12$ and $\gamma_{2, {\rm blue}} = 2.52\pm0.1$), the faint end slopes diverge significantly, with red quasars having a much steeper slope ($\gamma_{1, {\rm red}} = 0.62\pm0.20$ and $\gamma_{1, {\rm blue}} = 0.19\pm0.18$). 
With the large magnification of $\sim120$, the inferred luminosity of W2M~J1042+1641 may actually be probing the faint end slope, which could explain the high magnification bias and hence the high lens fraction found in this sample.

We note that the previously-discovered lensed red quasars are all radio sources. Some of them are intrinsically reddened, while others are likely reddened by the lensing galaxy itself.  
\citet{Malhotra97} showed that lensed quasars found in radio-selected samples have redder colors, implying that radio selection finds dusty systems that may be lost in optical quasar samples that often impose a blue color cut.  
Our surveying of red QSOs in shallow, wide-field surveys (i.e., W2M) is beginning to remedy this incompleteness by recovering radio-quiet reddened lensed QSOs. In addition, shallow flux-limited surveys benefit from increased magnification bias, making these lenses easier to find.  

Finding such a highly magnified red QSO offers a unique opportunity to study a population of QSOs whose luminosity is intrinsically lower than the QSOs found in wide-field surveys such as SDSS, which make up the vast majority of known QSOs.  

\section{Conclusions and future work} \label{sec:future}

We have discovered a quadruply lensed radio-quiet quasar at $z=2.517$ identified in through {\em HST} imaging of dust-reddened quasars selected by their {\em WISE} colors. Using optical and near-infrared spectroscopy, we determine that the quasar is reddened by $E(B-V) \simeq 0.7$ from dust intrinsic to the quasar's environment, and is not due to the lens.  Our lensing analysis find a magnification factor ranging between 53 and 122, depending on the model.

We estimate the quasar's black hole mass to be in the range $\log (M_{\rm BH} ~[M_\odot]) = 7.54 - 8.02$, depending on how the unreddened and demagnified continuum luminosity is computed.  The quasar's rest-frame infrared luminosity is $\log (L_{6\mu{\rm m}}~[{\rm erg~s^{-1}~Hz^{-1}}]) = 32.4 - 32.8$, which is near the knee of the QLF, representing more typical quasar luminosities that are difficult to access at high redshifts. 
The quasar's Eddington ratio could be as high as $L/L_{Edd} = 0.26$, if the bolometric luminosity is estimated from the 22\um\ flux, but could be as low as $L/L_{Edd} = 0.001$ if the {\em HST} F160W band is used.  The former would be consistent with accretion rates seen for more luminous red quasar samples.  We see evidence in the Einstein ring for enhanced star formation in the quasar's host and additional components that may suggest a galaxy merger, which is also consistent with the hosts of more luminous red quasars. 

On a short time scale, the main avenue for future work consists of further constraining the mass model by modeling the extended light distribution from the host galaxy, at the same time elucidating the nature of object X. This will be addressed in Rusu et al., in prep, which will also explore the position of W2M~J1042+1641 on the SMBH - host luminosity relation. The modeling residuals may further be improved by refining the PSF through an iterative approach \citep[e.g.,][]{Chen16}. Microlensing magnification maps may be explored to investigate the possibility of the flux anomalies being due to microlensing. On a longer timescale, monitoring observations would separate microlensing and intrinsic flux variations. Since the quasar is radio-faint, observations probing narrow line emission from the source, too spatially extended to be affected by microlensing \citep{Nierenberg14,Nierenberg17}, would be required to separate the effects of microlensing and substructure. 

\acknowledgments

This work was performed in part at Aspen Center for Physics, which is supported by National Science Foundation grant PHY-1607611.  EG thanks Crystal Martin for fruitful discussions during the Aspen workshop.
EG acknowledges the generous support of the Cottrell Scholar Award through the Research Corporation for Science Advancement. 
SGD and MJG acknowledge a partial support from the NSF grants AST-1413600 and AST-1518308, as well as the NASA grant 16-ADAP16-0232.

This work is based on GO observations made with the NASA/ESA Hubble Space Telescope from the Mikulski Archive for Space Telescopes (MAST), which is operated by the Association of Universities for Research in Astronomy, Inc., under NASA contract NAS5-26555. These observations are associated with program \#14706.

We thank the staff at the Keck observatory, where some of the data presented here were obtained. The authors recognize and acknowledge the very significant cultural role and reverence that the summit of Maunakea has always had within the indigenous Hawaiian community. We are most fortunate to have the opportunity to conduct observations from this mountain.

\vspace{5mm}
\facilities{HST(WFC3/IR), Keck(LRIS), IRTF(SpeX), LBT(MODS1B)}


\begin{thebibliography}{}
\bibitem[Angonin-Willaime et al.(1999)]{AngoninWillaime99} Angonin-Willaime, M.-C., Vanderriest, C., Courbin, F., et al.\ 1999, \aap, 347, 434 
\bibitem[Assef et al.(2011)]{Assef11} Assef, R.~J., Denney, K.~D., Kochanek, C.~S., et al.\ 2011, \apj, 742, 93 
\bibitem[Assef et al.(2013)]{Assef13} Assef, R.~J., Stern, D., Kochanek, C.~S., et al.\ 2013, \apj, 772, 26 
\bibitem[Assef et al.(2016)]{Assef16} Assef, R.~J., Walton, D.~J., Brightman, M., et al.\ 2016, \apj, 819, 111 
\bibitem[Assef et al.(2018)]{Assef18} Assef, R.~J., Stern, D., Noirot, G., et al.\ 2018, \apjs, 234, 23 
\bibitem[Banerji et al.(2012)]{Banerji12} Banerji, M., McMahon, R.~G., Hewett, P.~C., et al.\ 2012, \mnras, 427, 2275 
\bibitem[Becker et al.(1997)]{Becker97} Becker, R.~H., Gregg, M.~D., Hook, I.~M., et al.\ 1997, \apjl, 479, L93 
\bibitem[Bentz et al.(2004)]{Bentz04} Bentz, M.~C., Hall, P.~B., \& Osmer, P.~S.\ 2004, \aj, 128, 561 
\bibitem[Brusa et al.(2015)]{Brusa15} Brusa, M., Bongiorno, A., Cresci, G., et al.\ 2015, \mnras, 446, 2394 
\bibitem[Chen et al.(2016)]{Chen16} Chen, G.~C.-F., Suyu, S.~H., Wong, K.~C., et al.\ 2016, \mnras, 462, 3457 
\bibitem[Cushing et al.(2004)]{Cushing04} Cushing, M.~C., Vacca, W.~D., \& Rayner, J.~T.\ 2004, \pasp, 116, 362 
\bibitem[Curran et al.(2007)]{Curran07} Curran, S.~J., Darling, J., Bolatto, A.~D., et al.\ 2007, \mnras, 382, L11 
\bibitem[Di Matteo et al.(2005)]{DiMatteo05} Di Matteo, T., Springel, V., \& Hernquist, L.\ 2005, \nat, 433, 604 
\bibitem[Donevski et al.(2018)]{Donevski18} Donevski, D., Buat, V., Boone, F., et al.\ 2018, \aap, 614, A33 
\bibitem[Donley et al.(2012)]{Donley12} Donley, J.~L., Koekemoer, A.~M., Brusa, M., et al.\ 2012, \apj, 748, 142 
\bibitem[Faber \& Jackson(1976)]{faber76} Faber, S.~M., \& Jackson, R.~E.\ 1976, \apj, 204, 668 
\bibitem[Ferrarese \& Merritt(2000)]{Ferrarese00} Ferrarese, L., \& Merritt, D.\ 2000, \apjl, 539, L9 
\bibitem[Fosbury et al.(2003)]{Fosbury03} Fosbury, R.~A.~E., Villar-Mart{\'{\i}}n, M., Humphrey, A., et al.\ 2003, \apj, 596, 797 
\bibitem[Gavazzi et al.(2012)]{Gavazzi12} Gavazzi, R., Treu, T., Marshall, P.~J., Brault, F., \& Ruff, A.\ 2012, \apj, 761, 170 
\bibitem[Gebhardt et al.(2000)]{Gebhardt00} Gebhardt, K., Bender, R., Bower, G., et al.\ 2000, \apjl, 539, L13 
\bibitem[Glikman et al.(2004)]{Glikman04} Glikman, E., Gregg, M.~D., Lacy, M., et al.\ 2004, \apj, 607, 60 
\bibitem[Glikman et al.(2006)]{Glikman06} Glikman, E., Helfand, D.~J., \& White, R.~L.\ 2006, \apj, 640, 579 
\bibitem[Glikman et al.(2007a)]{Glikman07} Glikman, E., Helfand, D.~J., White, R.~L., et al.\ 2007, \apj, 667, 673 
\bibitem[Glikman et al.(2007b)]{Glikman07b} Glikman, E., Djorgovski, S.~G., Stern, D., Bogosavljevi{\'c}, M., \& Mahabal, A.\ 2007, \apjl, 663, L73 
\bibitem[Glikman et al.(2012)]{Glikman12} Glikman, E., Urrutia, T., Lacy, M., et al.\ 2012, \apj, 757, 51 
\bibitem[Glikman et al.(2013)]{Glikman13} Glikman, E., Urrutia, T., Lacy, M., et al.\ 2013, \apj, 778, 127 
\bibitem[Glikman et al.(2015)]{Glikman15} Glikman, E., Simmons, B., Mailly, M., et al.\ 2015, \apj, 806, 218 
\bibitem[Glikman et al.(2017)]{Glikman17} Glikman, E., LaMassa, S., Piconcelli, E., Urry, M., \& Lacy, M.\ 2017, \apj, 847, 116 
\bibitem[Glikman et al.(2018)]{Glikman18} Glikman, E., Lacy, M., LaMassa, S., et al.\ 2018, arXiv:1805.06961 
\bibitem[Gregg et al.(2002)]{Gregg02} Gregg, M.~D., Lacy, M., White, R.~L., et al.\ 2002, \apj, 564, 133 
\bibitem[Hall et al.(2002)]{Hall02} Hall, P.~B., Richards, G.~T., York, D.~G., et al.\ 2002, \apjl, 575, L51 
\bibitem[Hamann et al.(2017)]{Hamann17} Hamann, F., Zakamska, N.~L., Ross, N., et al.\ 2017, \mnras, 464, 3431 
\bibitem[Hewitt et al.(1992)]{Hewitt92} Hewitt, J.~N., Turner, E.~L., Lawrence, C.~R., Schneider, D.~P., \& Brody, J.~P.\ 1992, \aj, 104, 968 
\bibitem[Hopkins et al.(2005)]{Hopkins05} Hopkins, P.~F., Hernquist, L., Cox, T.~J., et al.\ 2005, \apj, 630, 705 
\bibitem[Hopkins \& Hernquist(2009)]{Hopkins09} Hopkins, P.~F., \& Hernquist, L.\ 2009, \apj, 698, 1550
\bibitem[Inada et al.(2003)]{Inada03} Inada, N., Becker, R.~H., Burles, S., et al.\ 2003, \aj, 126, 666  
\bibitem[Ivezi{\'c} et al.(2002)]{Ivezic02} Ivezi{\'c}, {\v Z}., Menou, K., Knapp, G.~R., et al.\ 2002, \aj, 124, 2364 
\bibitem[Keeton et al.(1998)]{Keeton98} Keeton, C.~R., Kochanek, C.~S., \& Falco, E.~E.\ 1998, \apj, 509, 561 
\bibitem[Keeton et al.(2003)]{Keeton03} Keeton, C.~R., Gaudi, B.~S., \& Petters, A.~O.\ 2003, \apj, 598, 138
\bibitem[Keeton et al.(2006)]{Keeton06} Keeton, C.~R., Burles, S., Schechter, P.~L., \& Wambsganss, J.\ 2006, \apj, 639, 1 
\bibitem[Kim et al.(2015)]{Kim15} Kim, D., Im, M., Glikman, E., Woo, J.-H., \& Urrutia, T.\ 2015, \apj, 812, 66 
\bibitem[Kochanek(1994)]{Kochanek94} Kochanek, C.~S.\ 1994, \apj, 436, 56 
\bibitem[Lacy et al.(2002)]{Lacy02} Lacy, M., Gregg, M., Becker, R.~H., et al.\ 2002, \aj, 123, 2925 
\bibitem[Lacy et al.(2004)]{Lacy04} Lacy, M., Storrie-Lombardi, L.~J., Sajina, A., et al.\ 2004, \apjs, 154, 166 
\bibitem[Lacy et al.(2015)]{Lacy15} Lacy, M., Ridgway, S.~E., Sajina, A., et al.\ 2015, \apj, 802, 102 
\bibitem[Larkin et al.(1994)]{Larkin94} Larkin, J.~E., Matthews, K., Lawrence, C.~R., et al.\ 1994, \apjl, 420, L9 
\bibitem[Lawrence et al.(1995)]{Lawrence95} Lawrence, C.~R., Elston, R., Januzzi, B.~T., \& Turner, E.~L.\ 1995, \aj, 110, 2570 
\bibitem[MacLeod et al.(2015)]{Macleod15} MacLeod, C.~L., Morgan, C.~W., Mosquera, A., et al.\ 2015, \apj, 806, 258 
\bibitem[Malhotra et al.(1997)]{Malhotra97} Malhotra, S., Rhoads, J.~E., \& Turner, E.~L.\ 1997, \mnras, 288, 138 
\bibitem[Negrello et al.(2007)]{Negrello07} Negrello, M., Perrotta, F., Gonz{\'a}lez-Nuevo, J., et al.\ 2007, \mnras, 377, 1557
\bibitem[Nierenberg et al.(2014)]{Nierenberg14} Nierenberg, A.~M., Treu, T., Wright, S.~A., Fassnacht, C.~D., \& Auger, M.~W.\ 2014, \mnras, 442, 2434 
\bibitem[Nierenberg et al.(2017)]{Nierenberg17} Nierenberg, A.~M., Treu, T., Brammer, G., et al.\ 2017, \mnras, 471, 2224 
\bibitem[Oguri(2010)]{Oguri10} Oguri, M.\ 2010, \pasj, 62, 1017 
\bibitem[Oguri \& Marshall(2010)]{OM10} Oguri, M., \& Marshall, P.~J.\ 2010, \mnras, 405, 2579 
\bibitem[Peng et al.(2002)]{Peng02} Peng, C.~Y., Ho, L.~C., Impey, C.~D., \& Rix, H.-W.\ 2002, \aj, 124, 266 
\bibitem[Peth et al.(2011)]{Peth11} Peth, M.~A., Ross, N.~P., \& Schneider, D.~P.\ 2011, \aj, 141, 105 
\bibitem[Rayner et al.(1998)]{Rayner98} Rayner, J.~T., Toomey, D.~W., Onaka, P.~M., et al.\ 1998, \procspie, 3354, 468 
\bibitem[Richards et al.(2006)]{Richards06} Richards, G.~T., Lacy, M., Storrie-Lombardi, L.~J., et al.\ 2006, \apjs, 166, 470 
\bibitem[Rusu et al.(2016)]{Rusu16} Rusu, C.~E., Oguri, M., Minowa, Y., et al.\ 2016, \mnras, 458, 2
\bibitem[Sanders et al.(1988)]{Sanders88} Sanders, D.~B., Soifer, B.~T., Elias, J.~H., et al.\ 1988, \apj, 325, 74 
\bibitem[Saxton et al.(2008)]{Saxton08} Saxton, R.~D., Read, A.~M., Esquej, P., et al.\ 2008, \aap, 480, 611 
\bibitem[Schechter \& Wambsganss(2002)]{Schechter02} Schechter, P.~L., \& Wambsganss, J.\ 2002, \apj, 580, 685 
\bibitem[Shen \& Liu(2012)]{Shen12} Shen, Y., \& Liu, X.\ 2012, \apj, 753, 125 
\bibitem[Sluse et al.(2012)]{Sluse12a} Sluse, D., Hutsem{\'e}kers, D., Courbin, F., Meylan, G., \& Wambsganss, J.\ 2012, \aap, 544, A62
\bibitem[Sluse et al.(2012)]{Sluse12b} Sluse, D., Chantry, V., Magain, P., Courbin, F., \& Meylan, G.\ 2012, \aap, 538, A99
\bibitem[Stern et al.(2005)]{Stern05} Stern, D., Eisenhardt, P., Gorjian, V., et al.\ 2005, \apj, 631, 163 
\bibitem[Stern et al.(2012)]{Stern12} Stern, D., Assef, R.~J., Benford, D.~J., et al.\ 2012, \apj, 753, 30 
\bibitem[Telfer et al.(2002)]{Telfer02} Telfer, R.~C., Zheng, W., Kriss, G.~A., \& Davidsen, A.~F.\ 2002, \apj, 565, 773 
\bibitem[Trump et al.(2006)]{Trump06} Trump, J.~R., Hall, P.~B., Reichard, T.~A., et al.\ 2006, \apjs, 165, 1 
\bibitem[Urrutia et al.(2008)]{Urrutia08} Urrutia, T., Lacy, M., \& Becker, R.~H.\ 2008, \apj, 674, 80-96 
\bibitem[Urrutia et al.(2009)]{Urrutia09} Urrutia, T., Becker, R.~H., White, R.~L., et al.\ 2009, \apj, 698, 1095 
\bibitem[Vacca et al.(2003)]{Vacca03} Vacca, W.~D., Cushing, M.~C., \& Rayner, J.~T.\ 2003, \pasp, 115, 389 
\bibitem[Vanden Berk et al.(2001)]{VandenBerk01} Vanden Berk, D.~E., Richards, G.~T., Bauer, A., et al.\ 2001, \aj, 122, 549 
\bibitem[Veilleux et al.(2013)]{Veilleux13} Veilleux, S., Trippe, M., Hamann, F., et al.\ 2013, \apj, 764, 15 
\bibitem[Warren et al.(2000)]{Warren00} Warren, S.~J., Hewett, P.~C., \& Foltz, C.~B.\ 2000, \mnras, 312, 827 
\bibitem[Winn et al.(2002)]{Winn02} Winn, J.~N., Lovell, J.~E.~J., Chen, H.-W., et al.\ 2002, \apj, 564, 143 
\end{thebibliography}
\end{document}